\documentclass{article}

\usepackage{arxiv}

\usepackage[utf8]{inputenc} 
\usepackage[T1]{fontenc}    
\usepackage{lmodern}        
\usepackage{hyperref}       
\usepackage{url}            
\usepackage{booktabs}       
\usepackage{amsfonts}       
\usepackage{nicefrac}       
\usepackage{microtype}      
\usepackage{graphicx}

\title{Replication of SARS-CoV-2 mutation analysis suggests differences
in per-protein mutation characteristics}

\author{
    William Svoboda
    \thanks{website: \url{https://will.cx}, email:
\href{mailto:wcsvoboda@gmail.com}{\nolinkurl{wcsvoboda@gmail.com}}}
   \\
    Computer Science \\
    Princeton University \\
  Princeton, NJ \\
  \texttt{\href{mailto:wsvoboda@princeton.edu}{\nolinkurl{wsvoboda@princeton.edu}}} \\
   \And
    Brendan McManamon
   \\
    Electrical and Computer Engineering \\
    Princeton University \\
  Princeton, NJ \\
  \texttt{\href{mailto:bm18@princeton.edu}{\nolinkurl{bm18@princeton.edu}}} \\
   \And
    Sara Schwartz
   \\
    Computer Science \\
    Princeton University \\
  Princeton, NJ \\
  \texttt{\href{mailto:sarats@Princeton.edu}{\nolinkurl{sarats@Princeton.edu}}} \\
  }

\usepackage{color}
\usepackage{fancyvrb}

\DefineVerbatimEnvironment{Highlighting}{Verbatim}{commandchars=\\\{\}}
\usepackage{framed}
\definecolor{shadecolor}{RGB}{248,248,248}
\newenvironment{Shaded}{\begin{snugshade}}{\end{snugshade}}

\newcommand{\AttributeTok}[1]{\textcolor[rgb]{0.77,0.63,0.00}{#1}}

\newcommand{\CommentTok}[1]{\textcolor[rgb]{0.56,0.35,0.01}{\textit{#1}}}

\newcommand{\ConstantTok}[1]{\textcolor[rgb]{0.00,0.00,0.00}{#1}}
\newcommand{\ControlFlowTok}[1]{\textcolor[rgb]{0.13,0.29,0.53}{\textbf{#1}}}

\newcommand{\DecValTok}[1]{\textcolor[rgb]{0.00,0.00,0.81}{#1}}

\newcommand{\FloatTok}[1]{\textcolor[rgb]{0.00,0.00,0.81}{#1}}
\newcommand{\FunctionTok}[1]{\textcolor[rgb]{0.00,0.00,0.00}{#1}}

\newcommand{\NormalTok}[1]{#1}

\newcommand{\OtherTok}[1]{\textcolor[rgb]{0.56,0.35,0.01}{#1}}

\newcommand{\SpecialCharTok}[1]{\textcolor[rgb]{0.00,0.00,0.00}{#1}}

\newcommand{\StringTok}[1]{\textcolor[rgb]{0.31,0.60,0.02}{#1}}

\newlength{\csllabelwidth}
\newlength{\cslhangindent}
  {}%
  {\par}
\newenvironment{CSLReferences}[2] 
 {
  \setlength{\parindent}{0pt}
  \ifodd #1 \everypar{\setlength{\hangindent}{\cslhangindent}}\ignorespaces\fi
  \ifnum #2 > 0
  \setlength{\parskip}{#2\baselineskip}
  \fi
 }%
 {}
\usepackage{calc} 

\newcommand{\CSLLeftMargin}[1]{\parbox[t]{\csllabelwidth}{#1}}
\newcommand{\CSLRightInline}[1]{\parbox[t]{\linewidth - \csllabelwidth}{#1}\break}

\usepackage{fvextra}
\DefineVerbatimEnvironment{Highlighting}{Verbatim}{breaklines,commandchars=\\\{\}}
\hypersetup{colorlinks=true,linkcolor=blue,citecolor=black,urlcolor=black}
\usepackage{booktabs}

\begin{document}
\maketitle

\def\tightlist{}

\begin{abstract}
The increasing spread of COVID-19, caused by the virus SARS-CoV-2,
raises concerns about the extent to which mutations have occurred across
the viral genome. We present a partial replication of an earlier 2021
study by Wang, R. \emph{et al.} that determined the presence of four
substrains and eleven top mutations in the United States. We analyze a
portion of the authors' data set in order to recreate Figure S1 from the
paper, recapitulating the same features observed in the original figure.
We further generate a summary of mutation characteristics for each of
the 26 named proteins and confirm the significance of the spike protein
at roughly 24\% of all recorded mutations. Our analysis suggests that
additional factors may affect per-protein mutation rate besides protein
length.
\end{abstract}


\hypersetup{urlcolor=blue}

\hypertarget{background}{%
\section{Background}\label{background}}

The spread of coronavirus disease 2019 (COVID-19) continues to be of
scientific and public health interest in the context of the ongoing
global pandemic. Of particular concern is whether the underlying virus,
severe acute respiratory syndrome coronavirus 2 (SARS-CoV-2), has become
more infectious as a result of new mutations.

A 2021 paper by Wang, R. \emph{et
al.}\textsuperscript{\protect\hyperlink{ref-Wang2021}{1}} analyzed
45,494 complete SARS-CoV-2 genome sequences in order to analyze their
mutations. Of the 12,754 sequences from the United States, the authors
determined the presence of four substrains and eleven top mutations.
Furthermore, the study concluded that two of these substrains had become
more infectious given the mutations found on the spike protein.

\begin{figure}

{\centering \includegraphics[width=\textwidth]{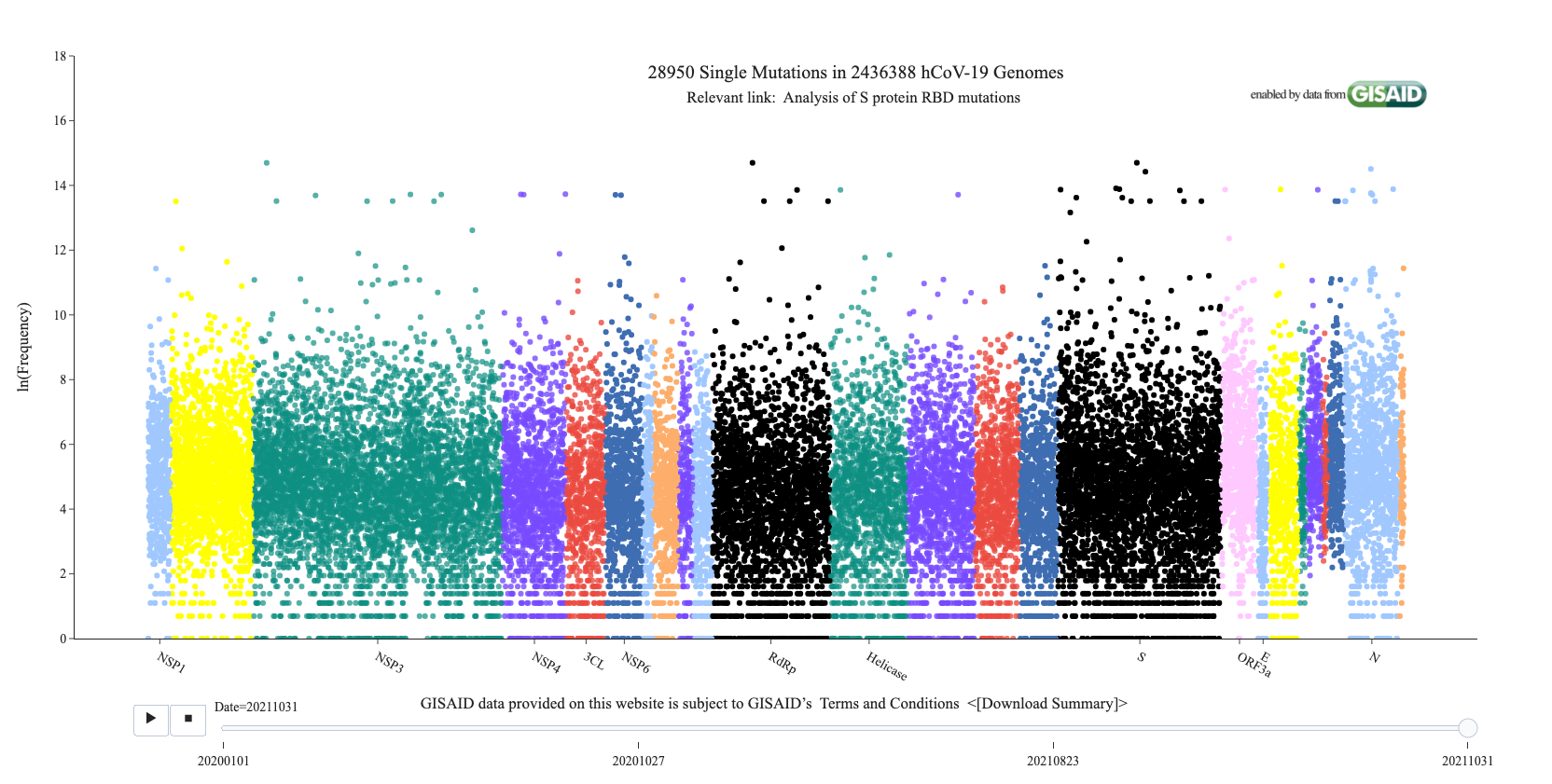} 

}

\caption{Screenshot of the mutation tracker website showing the distribution of mutations across the entire SARS-CoV-2 genome.}\label{fig:fig1}
\end{figure}

We present a partial replication of the original study in order to
validate its findings. A subset of the data used in the paper was
processed in order to recreate selected figures (\autoref{fig:fig1}).
Further analysis was also performed in order to summarize mutation
characteristics for each protein across the entire SARS-CoV-2 genome.
Finally, we discuss the implications of these findings on our current
understanding of the SARS-CoV-2 viral makeup.

\hypertarget{results}{%
\section{Results}\label{results}}

\hypertarget{mutation-tracker}{%
\subsection{Mutation Tracker}\label{mutation-tracker}}

\begin{figure}

{\centering \includegraphics[width=\textwidth]{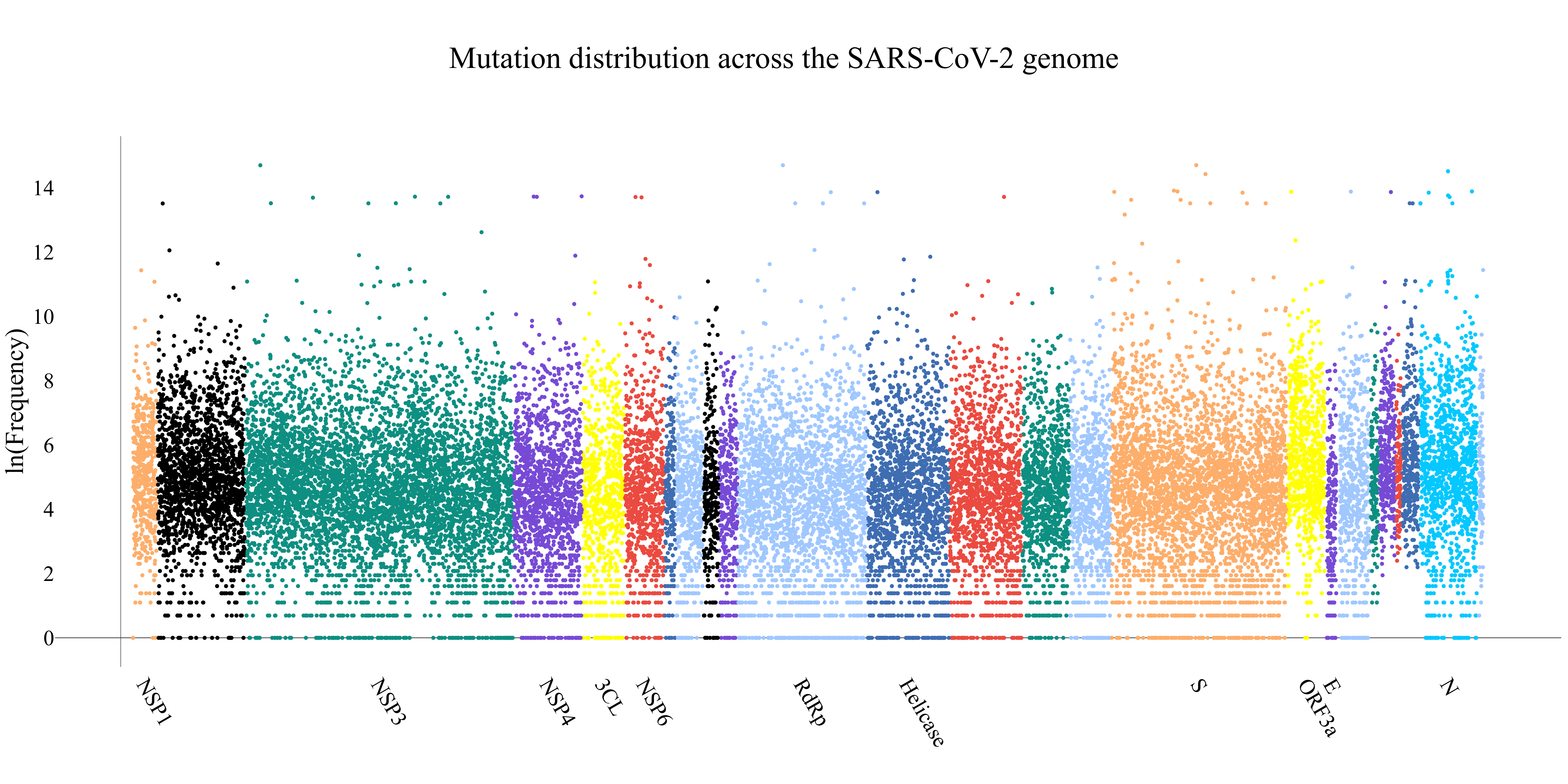} 

}

\caption{Genome-wide SARS-CoV-2 mutation distribution as of 31 October 2021. As in the original distribution figure, the natural log frequency of mutations on specific nucleotide positions is plotted on the $y$-axis. The positions of certain landmark proteins are also annotated on the $x$-axis. (author W.S.)}\label{fig:fig2}
\end{figure}

As part of the original
study\textsuperscript{\protect\hyperlink{ref-Wang2021}{1}}, a website
called
\href{https://users.math.msu.edu/users/weig/SARS-CoV-2_Mutation_Tracker.html}{Mutation
Tracker} was created to report the distribution of unique single
mutations across the entire SARS-CoV-2 genome. A screen shot of this
tracker is included in the
paper's\textsuperscript{\protect\hyperlink{ref-Wang2021}{1}}
supplementary information (Figure S1, included as \autoref{fig:fig1} in
this report). We downloaded the summary data provided from the website,
itself generated by Wang, R. \emph{et al.} using sequence data from
GISAID (\url{https://www.gisaid.org/}). The summary data is updated
automatically by the mutation website, with the totals from 31 October
2021 being used in this paper. Split among 26 different files for each
named protein, the summary data was reprocessed into a single data frame
representing the whole viral genome.

As \autoref{fig:fig2} shows, we replicated the original distribution
graph using this updated data. It is evident that NSP3 and the spike
protein are both among the largest proteins and have experienced a high
frequency of unique mutations across their domains. We also observe that
ORF7a, ORF7b, and ORF8 all have relatively high mutation frequencies, in
spite of their short length.

\hypertarget{mutation-summary-statistics}{%
\subsection{Mutation Summary
Statistics}\label{mutation-summary-statistics}}

\begin{table}

\caption{\label{tab:tab1}Summary statistics for each SARS-CoV-2 protein (author W.S.)}
\centering
\begin{tabular}[t]{lrrrr}
\toprule
protein & total mutations & \% total mutations & protein length & mutations/nucleotide\\
\midrule
Spike & 17,134,885 & 23.785 & 3,821 & 4,484.398\\
NSP3 & 12,191,479 & 16.923 & 5,834 & 2,089.729\\
Nucleocapsid & 11,760,145 & 16.324 & 1,259 & 9,340.862\\
RdRp & 7,193,224 & 9.985 & 2,794 & 2,574.525\\
NSP4 & 3,612,766 & 5.015 & 1,499 & 2,410.117\\
\addlinespace
NSP6 & 2,735,245 & 3.797 & 869 & 3,147.578\\
ORF3a & 2,490,822 & 3.457 & 827 & 3,011.877\\
NSP2 & 2,392,844 & 3.321 & 1,913 & 1,250.833\\
Helicase & 2,319,519 & 3.220 & 1,802 & 1,287.191\\
ORF8 & 2,072,786 & 2.877 & 365 & 5,678.866\\
\addlinespace
Exonuclease & 1,851,994 & 2.571 & 1,580 & 1,172.148\\
Membrane & 1,606,101 & 2.229 & 668 & 2,404.343\\
ORF7a & 1,462,425 & 2.030 & 365 & 4,006.644\\
2’-O-ribose MTases & 564,099 & 0.783 & 893 & 631.690\\
endoRNAse & 544,308 & 0.756 & 1,037 & 524.887\\
\addlinespace
3CL & 484,768 & 0.673 & 917 & 528.646\\
NSP1 & 433,783 & 0.602 & 539 & 804.792\\
NSP9 & 302,133 & 0.419 & 338 & 893.885\\
NSP8 & 254,185 & 0.353 & 593 & 428.642\\
ORF10 & 163,272 & 0.227 & 116 & 1,407.517\\
\addlinespace
NSP7 & 125,480 & 0.174 & 248 & 505.968\\
NSP10 & 122,023 & 0.169 & 416 & 293.325\\
ORF6 & 109,947 & 0.153 & 185 & 594.308\\
ORF7b & 56,865 & 0.079 & 131 & 434.084\\
Envelope & 53,802 & 0.075 & 227 & 237.013\\
\addlinespace
NSP11 & 2,578 & 0.004 & 38 & 67.842\\
\bottomrule
\end{tabular}
\end{table}

To quantify our interpretations, we computed further statistics to
summarize the mutation distribution. \autoref{tab:tab1} lists this data
for each protein ordered by total mutation frequency. The spike protein,
NSP3, and the nucleocapsid are the top three proteins, with their
respective frequencies (17,134,885, 12,191,479, and 11,760,145
mutations) being a full order of magnitude greater than the rest.
Interestingly, we note that ORF8, ORF7a, and NSP6 have an unusually high
number of mutations with respect to their length. This corroborates our
previous observations of the mutation distribution graph provided by
\autoref{fig:fig2}.

\hypertarget{mutation-percentage-comparison}{%
\subsection{Mutation Percentage
Comparison}\label{mutation-percentage-comparison}}

\begin{figure}

{\centering \includegraphics[width=\textwidth]{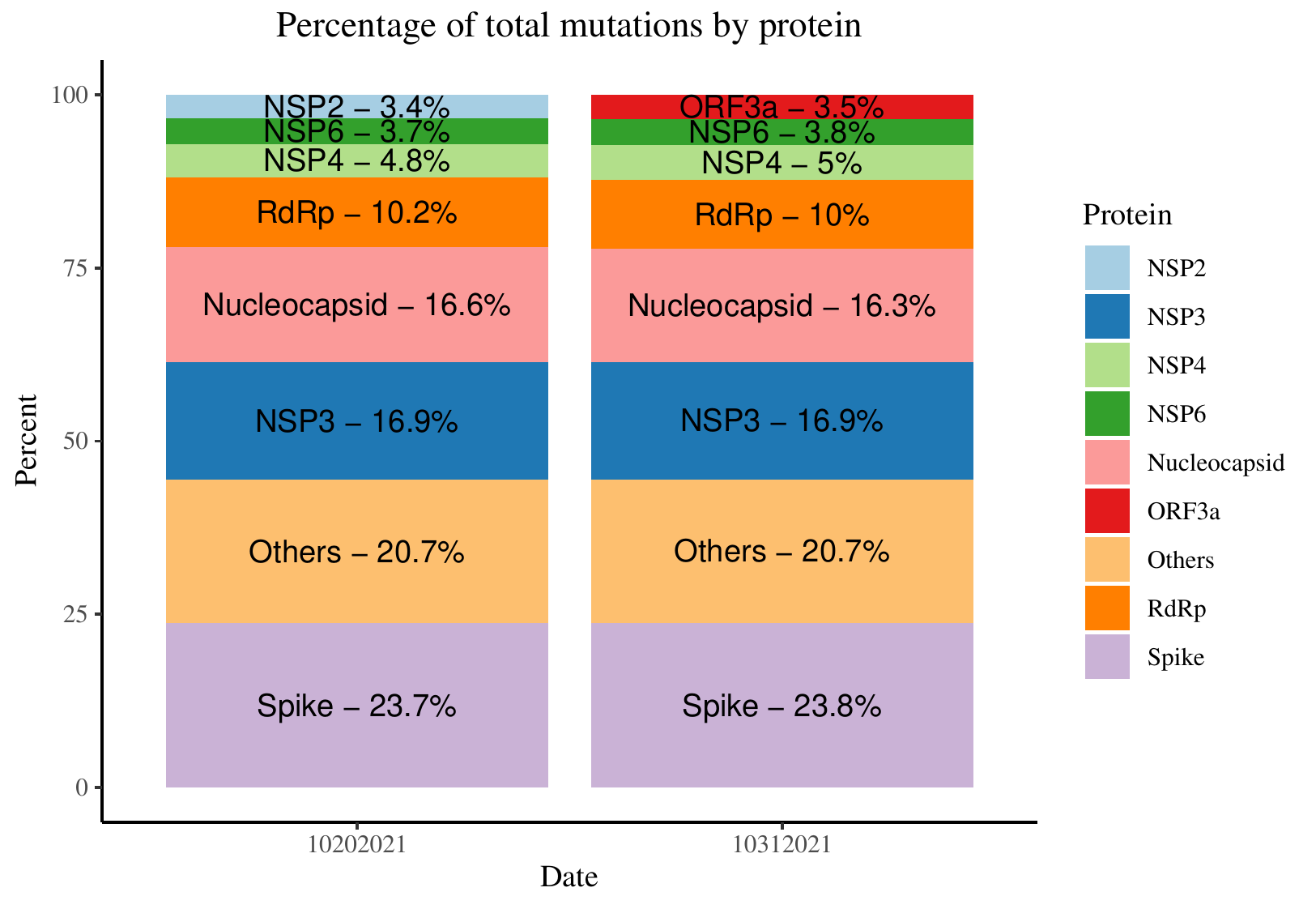} 

}

\caption{Comparison of total mutation percentages for two collected dates (listed in \texttt{MMDDYYYY} format). For each bar, only the 7 proteins with the largest percentages are named. (author W.S.)}\label{fig:fig3}
\end{figure}

From our summary statistics, we generated a graphical breakdown of the
total mutation percentage. \autoref{fig:fig3} displays this data for the
top 7 proteins between 20 October 2021 and 31 October 2021. Following
from our analysis of the summary statistics, the spike protein made up
roughly 24\% of the total recorded mutation frequency. The graph changes
little between the 10 days comprising each time point. Of note is that a
full 13 proteins make up less than 1\% each of the total.

\hypertarget{discussion}{%
\section{Discussion}\label{discussion}}

Our analysis successfully replicated the original plot of the mutation
distribution represented by \autoref{fig:fig1}. Our version,
\autoref{fig:fig2}, makes use of the updated data provided from the
Mutation Tracker website and recapitulates the features observed there.
Along with the summary statistics from \autoref{tab:tab1}, we
corroborated the significance of the spike protein in the current
pattern of SARS-CoV-2 mutation.

Our approach is limited by the scope of the data used in the study. The
Mutation Tracker website only allows downloading summary data from the
most recent time point. While we had preserved an earlier archive from
20 October 2021, a more thorough comparison than that seen in
\autoref{fig:fig3} would use a greater range of dates. Further study
into the change in mutation over time, either by sampling the tracker or
by directly processing the source data from GISAID, could provide more
insight.

The summary statistics from \autoref{tab:tab1} suggest differences in
mutation characteristics between the 26 studied proteins. Given our
observation of relatively more mutations in shorter proteins, a
conformational analysis could further assess the influence of a
protein's size and 3D structure on the number of mutations it
experiences.

There is also interest in determining the extent, if any, of
evolutionary pressure on mutation rate. Some of the proteins comprising
the lowest percentages of the total mutation frequency, such as NSP11
and the envelope protein, are known to have important roles in the viral
life
cycle\textsuperscript{\protect\hyperlink{ref-Zhang2016}{2},\protect\hyperlink{ref-Schoeman2019}{3}}.
Future work could determine if proteins with functions in viral
replication versus internal regulation experience more mutations.

\hypertarget{conclusion}{%
\section{Conclusion}\label{conclusion}}

In this paper, we presented a partial replication of an earlier study by
Wang, R. \emph{et
al.}\textsuperscript{\protect\hyperlink{ref-Wang2021}{1}}. We
recapitulated their visualization of the genome-wide SARS-CoV-2 mutation
distribution. We further analyzed the underlying data in order to
summarize the influence of each protein on the overall mutation
frequencies. The current global pandemic remains one of the most
significant challenges to public health in recent memory. In light of
emerging variants of SARS-CoV-2, it is even more important to understand
how and where the virus mutates.

\hypertarget{methods}{%
\section{Methods}\label{methods}}

\hypertarget{data-collection-and-preprocessing}{%
\subsection{Data Collection and
Preprocessing}\label{data-collection-and-preprocessing}}

The original paper\textsuperscript{\protect\hyperlink{ref-Wang2021}{1}}
used 45,494 complete genome sequences of SARS-CoV-2 strains from
infected invdividuals. These sequences were sourced from the GISAID
(\url{https://www.gisaid.org/}) database up to 11 September 2020. For
this analysis, we used the 31 October 2021 summary data from the
\href{https://users.math.msu.edu/users/weig/SARS-CoV-2_Mutation_Tracker.html}{Mutation
Tracker} website. The tracker is continually updated with the latest
data from GISAID.

Data preprocessing was accomplished using
R\textsuperscript{\protect\hyperlink{ref-R-base}{4}} and the
dplyr\textsuperscript{\protect\hyperlink{ref-R-dplyr}{5}} package. The
26 \texttt{.csv} files for each named protein were parsed and combined
into a single data frame in order of position on the SARS-CoV-2 genome.
Mutation frequencies were mapped to the corresponding protein name using
nucleotide positions sourced from the NIH
(\url{https://www.ncbi.nlm.nih.gov/gene/}).

\hypertarget{plots-and-figures}{%
\subsection{Plots and Figures}\label{plots-and-figures}}

The plot of the genome-wide SARS-CoV-2 mutation distribution
(\autoref{fig:fig2}) was generated using the Plotly package for
R\textsuperscript{\protect\hyperlink{ref-R-plotly}{6}}. The stacked bar
chart comparing per-protein mutation percentages (\autoref{fig:fig3})
was generated using the ggplot2
package\textsuperscript{\protect\hyperlink{ref-R-ggplot2}{7}} using the
same data as the mutation distribution plot (summarized for each
protein).

\hypertarget{declarations}{%
\section*{Declarations}\label{declarations}}
\addcontentsline{toc}{section}{Declarations}

\hypertarget{data-and-code-availability}{%
\subsection*{Data and Code
Availability}\label{data-and-code-availability}}
\addcontentsline{toc}{subsection}{Data and Code Availability}

This paper uses the summary data from the
\href{https://users.math.msu.edu/users/weig/SARS-CoV-2_Mutation_Tracker.html}{Mutation
Tracker} website up to 31 October 2021. This data was originally sourced
from GISAID (\url{https://www.gisaid.org/}).

All code for this paper is attached in the appendix and is also
available on the author's (W.S.) GitHub at
\url{https://github.com/thisstillwill/QCB455-Fall-2021}.

\hypertarget{acknowledgements}{%
\subsection*{Acknowledgements}\label{acknowledgements}}
\addcontentsline{toc}{subsection}{Acknowledgements}

The authors are grateful to Rui Wang and Guo-Wei Wei for their
correspondance providing the color values used in the Mutation Tracker
and describing the approach used to clean the summary data.

\hypertarget{author-contributions}{%
\subsection*{Author Contributions}\label{author-contributions}}
\addcontentsline{toc}{subsection}{Author Contributions}

W.S. downloaded and preprocessed the summary data from the Mutation
Tracker website. W.S. plotted the mutation frequencies from the summary
data to replicate Figure S1 from the original paper. W.S. conducted
further analysis of the dataset to summarize the mutation data for each
protein.

B.M. downloaded and translated reference genomes for related
coronaviruses into individual proteins. B.M. conducted multiple sequence
alignment for each protein and calculated percent similarity in order to
replicate Figures S5, S7, S8, S10, and S13. Additionally, B.M. added a
table summarizing substitution mutation frequencies.

S.S. conducted multiple sequence alignment and single-nucleotide
polymorphism (SNP) calling on the complete genome sequences to create
SNP profiles for each genome, and used a Jaccard distance-based
representation of the SNP variants as input features into a K-means
clustering algorithm. S.S. plotted the distributions of the resulting
clusters on a U.S. map to replicate Figure 1 from the original paper.

All authors have read and agreed to the published version of the
manuscript.

\hypertarget{competing-interests}{%
\subsection*{Competing Interests}\label{competing-interests}}
\addcontentsline{toc}{subsection}{Competing Interests}

The authors declare no competing interests.

\hypertarget{references}{%
\section*{References}\label{references}}
\addcontentsline{toc}{section}{References}

\hypertarget{refs}{}
\begin{CSLReferences}{0}{0}
\leavevmode\vadjust pre{\hypertarget{ref-Wang2021}{}}%
\CSLLeftMargin{1. }
\CSLRightInline{Wang, R. \emph{et al.}
\href{https://doi.org/10.1038/s42003-021-01754-6}{Analysis of SARS-CoV-2
mutations in the united states suggests presence of four substrains and
novel variants}. \emph{Communications Biology} \textbf{4}, 228 (2021).}

\leavevmode\vadjust pre{\hypertarget{ref-Zhang2016}{}}%
\CSLLeftMargin{2. }
\CSLRightInline{Zhang, M. \emph{et al.}
\href{https://doi.org/10.1128/JVI.01309-16}{Structural biology of the
arterivirus nsp11 endoribonucleases}. \emph{Journal of virology}
\textbf{91}, e01309--16 (2016).}

\leavevmode\vadjust pre{\hypertarget{ref-Schoeman2019}{}}%
\CSLLeftMargin{3. }
\CSLRightInline{Schoeman, D. \& Fielding, B. C.
\href{https://doi.org/10.1186/s12985-019-1182-0}{Coronavirus envelope
protein: Current knowledge}. \emph{Virology Journal} \textbf{16}, 69
(2019).}

\leavevmode\vadjust pre{\hypertarget{ref-R-base}{}}%
\CSLLeftMargin{4. }
\CSLRightInline{R Core Team. \emph{\href{https://www.R-project.org/}{R:
A language and environment for statistical computing}}. (R Foundation
for Statistical Computing, 2021).}

\leavevmode\vadjust pre{\hypertarget{ref-R-dplyr}{}}%
\CSLLeftMargin{5. }
\CSLRightInline{Wickham, H., François, R., Henry, L. \& Müller, K.
\emph{\href{https://CRAN.R-project.org/package=dplyr}{Dplyr: A grammar
of data manipulation}}. (2021).}

\leavevmode\vadjust pre{\hypertarget{ref-R-plotly}{}}%
\CSLLeftMargin{6. }
\CSLRightInline{Sievert, C. \emph{et al.}
\emph{\href{https://CRAN.R-project.org/package=plotly}{Plotly: Create
interactive web graphics via plotly.js}}. (2021).}

\leavevmode\vadjust pre{\hypertarget{ref-R-ggplot2}{}}%
\CSLLeftMargin{7. }
\CSLRightInline{Wickham, H. \emph{et al.}
\emph{\href{https://CRAN.R-project.org/package=ggplot2}{ggplot2: Create
elegant data visualisations using the grammar of graphics}}. (2021).}

\leavevmode\vadjust pre{\hypertarget{ref-R-bookdown}{}}%
\CSLLeftMargin{8. }
\CSLRightInline{Xie, Y.
\emph{\href{https://CRAN.R-project.org/package=bookdown}{Bookdown:
Authoring books and technical documents with r markdown}}. (2021).}

\leavevmode\vadjust pre{\hypertarget{ref-R-formatR}{}}%
\CSLLeftMargin{9. }
\CSLRightInline{Xie, Y.
\emph{\href{https://github.com/yihui/formatR}{formatR: Format r code
automatically}}. (2021).}

\leavevmode\vadjust pre{\hypertarget{ref-R-knitr}{}}%
\CSLLeftMargin{10. }
\CSLRightInline{Xie, Y. \emph{\href{https://yihui.org/knitr/}{Knitr: A
general-purpose package for dynamic report generation in r}}. (2021).}

\leavevmode\vadjust pre{\hypertarget{ref-R-reticulate}{}}%
\CSLLeftMargin{11. }
\CSLRightInline{Ushey, K., Allaire, J. \& Tang, Y.
\emph{\href{https://github.com/rstudio/reticulate}{Reticulate: Interface
to python}}. (2021).}

\leavevmode\vadjust pre{\hypertarget{ref-R-rticles}{}}%
\CSLLeftMargin{12. }
\CSLRightInline{Allaire, J. \emph{et al.}
\emph{\href{https://github.com/rstudio/rticles}{Rticles: Article formats
for r markdown}}. (2021).}

\leavevmode\vadjust pre{\hypertarget{ref-bookdown2016}{}}%
\CSLLeftMargin{13. }
\CSLRightInline{Xie, Y.
\emph{\href{https://bookdown.org/yihui/bookdown}{Bookdown: Authoring
books and technical documents with {R} markdown}}. (Chapman; Hall/CRC,
2016).}

\leavevmode\vadjust pre{\hypertarget{ref-ggplot22016}{}}%
\CSLLeftMargin{14. }
\CSLRightInline{Wickham, H.
\emph{\href{https://ggplot2.tidyverse.org}{ggplot2: Elegant graphics for
data analysis}}. (Springer-Verlag New York, 2016).}

\leavevmode\vadjust pre{\hypertarget{ref-knitr2015}{}}%
\CSLLeftMargin{15. }
\CSLRightInline{Xie, Y. \emph{\href{https://yihui.org/knitr/}{Dynamic
documents with {R} and knitr}}. (Chapman; Hall/CRC, 2015).}

\leavevmode\vadjust pre{\hypertarget{ref-knitr2014}{}}%
\CSLLeftMargin{16. }
\CSLRightInline{Xie, Y.
\href{http://www.crcpress.com/product/isbn/9781466561595}{Knitr: A
comprehensive tool for reproducible research in {R}}. in
\emph{Implementing reproducible computational research} (eds. Stodden,
V., Leisch, F. \& Peng, R. D.) (Chapman; Hall/CRC, 2014).}

\leavevmode\vadjust pre{\hypertarget{ref-plotly2020}{}}%
\CSLLeftMargin{17. }
\CSLRightInline{Sievert, C.
\emph{\href{https://plotly-r.com}{Interactive web-based data
visualization with r, plotly, and shiny}}. (Chapman; Hall/CRC, 2020).}

\end{CSLReferences}

\appendix

\hypertarget{all-code}{%
\section{All Code}\label{all-code}}

\begin{Shaded}
\begin{Highlighting}[]
\CommentTok{\# Required packages}
\FunctionTok{library}\NormalTok{(knitr, }\AttributeTok{quietly =} \ConstantTok{TRUE}\NormalTok{)}
\FunctionTok{library}\NormalTok{(formatR, }\AttributeTok{quietly =} \ConstantTok{TRUE}\NormalTok{)}
\FunctionTok{library}\NormalTok{(rticles, }\AttributeTok{quietly =} \ConstantTok{TRUE}\NormalTok{)}
\FunctionTok{library}\NormalTok{(dplyr, }\AttributeTok{quietly =} \ConstantTok{TRUE}\NormalTok{)}
\FunctionTok{library}\NormalTok{(reticulate, }\AttributeTok{quietly =} \ConstantTok{TRUE}\NormalTok{)}
\FunctionTok{library}\NormalTok{(plotly, }\AttributeTok{quietly =} \ConstantTok{TRUE}\NormalTok{)}
\FunctionTok{library}\NormalTok{(bookdown, }\AttributeTok{quietly =} \ConstantTok{TRUE}\NormalTok{)}

\CommentTok{\# Wrap R code in final document}
\NormalTok{knitr}\SpecialCharTok{::}\NormalTok{opts\_chunk}\SpecialCharTok{$}\FunctionTok{set}\NormalTok{(}\AttributeTok{cache =} \ConstantTok{TRUE}\NormalTok{, }\AttributeTok{echo =} \ConstantTok{FALSE}\NormalTok{, }\AttributeTok{message =} \ConstantTok{FALSE}\NormalTok{, }\AttributeTok{error =} \ConstantTok{FALSE}\NormalTok{,}
    \AttributeTok{fig.align =} \StringTok{"center"}\NormalTok{, }\AttributeTok{out.width =} \StringTok{"}\SpecialCharTok{\textbackslash{}\textbackslash{}}\StringTok{textwidth"}\NormalTok{, }\AttributeTok{tidy.opts =} \FunctionTok{list}\NormalTok{(}\AttributeTok{width.cutoff =} \DecValTok{80}\NormalTok{),}
    \AttributeTok{tidy =} \ConstantTok{TRUE}\NormalTok{)}
\NormalTok{knitr}\SpecialCharTok{::}\FunctionTok{include\_graphics}\NormalTok{(}\StringTok{"./figures/original\_distribution.png"}\NormalTok{)}
\CommentTok{\# Read latest mutation summary data into a combined data frame Data retrieved}
\CommentTok{\# from https://users.math.msu.edu/users/weig/SARS{-}CoV{-}2\_Mutation\_Tracker.html}
\NormalTok{path }\OtherTok{\textless{}{-}} \StringTok{"./data/MutationSummary\_10312021/"}
\NormalTok{files }\OtherTok{\textless{}{-}} \FunctionTok{dir}\NormalTok{(path)}
\NormalTok{mutation\_summaries }\OtherTok{\textless{}{-}} \FunctionTok{do.call}\NormalTok{(rbind, }\FunctionTok{lapply}\NormalTok{(}\FunctionTok{paste0}\NormalTok{(path, files), read.csv))}

\CommentTok{\# Add column for raw nucleotide position}
\NormalTok{mutation\_summaries }\OtherTok{\textless{}{-}}\NormalTok{ mutation\_summaries }\SpecialCharTok{\%\textgreater{}\%}
    \FunctionTok{mutate}\NormalTok{(}\AttributeTok{position =} \FunctionTok{as.integer}\NormalTok{(}\FunctionTok{gsub}\NormalTok{(}\StringTok{"[\^{}0{-}9]"}\NormalTok{, }\StringTok{""}\NormalTok{, mutation\_site))) }\SpecialCharTok{\%\textgreater{}\%}
    \FunctionTok{arrange}\NormalTok{(position)}

\CommentTok{\# Create new data frame to track the combined frequencies of each unique}
\CommentTok{\# mutation}
\NormalTok{mutation\_summaries\_combined }\OtherTok{\textless{}{-}}\NormalTok{ mutation\_summaries }\SpecialCharTok{\%\textgreater{}\%}
    \FunctionTok{group\_by}\NormalTok{(position) }\SpecialCharTok{\%\textgreater{}\%}
    \FunctionTok{summarise}\NormalTok{(}\AttributeTok{frequency =} \FunctionTok{sum}\NormalTok{(Total\_frequency))}

\CommentTok{\# Use nucleotide locations from https://www.ncbi.nlm.nih.gov/gene/ to clean}
\CommentTok{\# data Protein name mapping adapted from original paper (Python code): def}
\CommentTok{\# mapDict(self): keys = [\textquotesingle{}[266:805]\textquotesingle{}, \textquotesingle{}[806:2719]\textquotesingle{}, \textquotesingle{}[2720:8554]\textquotesingle{},}
\CommentTok{\# \textquotesingle{}[8555:10054]\textquotesingle{}, \textquotesingle{}[10055:10972]\textquotesingle{}, \textquotesingle{}[10973:11842]\textquotesingle{}, \textquotesingle{}[11843:12091]\textquotesingle{},}
\CommentTok{\# \textquotesingle{}[12092:12685]\textquotesingle{}, \textquotesingle{}[12686:13024]\textquotesingle{}, \textquotesingle{}[13025:13441]\textquotesingle{}, \textquotesingle{}[13442:16236]\textquotesingle{},}
\CommentTok{\# \textquotesingle{}[16237:18039]\textquotesingle{}, \textquotesingle{}[18040:19620]\textquotesingle{}, \textquotesingle{}[19621:20658]\textquotesingle{}, \textquotesingle{}[20659:21552]\textquotesingle{},}
\CommentTok{\# \textquotesingle{}[13442:13480]\textquotesingle{}, \textquotesingle{}[21563:25384]\textquotesingle{}, \textquotesingle{}[25393:26220]\textquotesingle{}, \textquotesingle{}[26245:26472]\textquotesingle{},}
\CommentTok{\# \textquotesingle{}[26523:27191]\textquotesingle{}, \textquotesingle{}[27202:27387]\textquotesingle{}, \textquotesingle{}[27394:27759]\textquotesingle{}, \textquotesingle{}[27756:27887]\textquotesingle{},}
\CommentTok{\# \textquotesingle{}[27894:28259]\textquotesingle{}, \textquotesingle{}[28274:29533]\textquotesingle{}, \textquotesingle{}29558:29674\textquotesingle{}] values = [\textquotesingle{}NSP1\textquotesingle{}, \textquotesingle{}NSP2\textquotesingle{},}
\CommentTok{\# \textquotesingle{}NSP3\textquotesingle{}, \textquotesingle{}NSP4\textquotesingle{}, \textquotesingle{}3CL\textquotesingle{}, \textquotesingle{}NSP6\textquotesingle{}, \textquotesingle{}NSP7\textquotesingle{}, \textquotesingle{}NSP8\textquotesingle{}, \textquotesingle{}NSP9\textquotesingle{}, \textquotesingle{}NSP10\textquotesingle{}, \textquotesingle{}RdRp\textquotesingle{},}
\CommentTok{\# \textquotesingle{}Helicase\textquotesingle{}, \textquotesingle{}Exonuclease\textquotesingle{}, \textquotesingle{}endoRNAse\textquotesingle{}, \textquotesingle{}2’{-}O{-}ribose MTases\textquotesingle{},}
\CommentTok{\# \textquotesingle{}NSP11\textquotesingle{},\textquotesingle{}Spike\textquotesingle{}, \textquotesingle{}ORF3a\textquotesingle{}, \textquotesingle{}Envelope\textquotesingle{}, \textquotesingle{}Membrane\textquotesingle{}, \textquotesingle{}ORF6\textquotesingle{}, \textquotesingle{}ORF7a\textquotesingle{}, \textquotesingle{}ORF7b\textquotesingle{},}
\CommentTok{\# \textquotesingle{}ORF8\textquotesingle{}, \textquotesingle{}Nucleocapsid\textquotesingle{}, \textquotesingle{}ORF10\textquotesingle{}] nameMap = dict(zip(keys,values)) return}
\CommentTok{\# nameMap}
\NormalTok{name\_map }\OtherTok{\textless{}{-}} \FunctionTok{rep}\NormalTok{(}\StringTok{"Others"}\NormalTok{, }\DecValTok{30000}\NormalTok{)}
\NormalTok{name\_map[}\DecValTok{266}\SpecialCharTok{:}\DecValTok{805}\NormalTok{] }\OtherTok{\textless{}{-}} \StringTok{"NSP1"}
\NormalTok{name\_map[}\DecValTok{806}\SpecialCharTok{:}\DecValTok{2719}\NormalTok{] }\OtherTok{\textless{}{-}} \StringTok{"NSP2"}
\NormalTok{name\_map[}\DecValTok{2720}\SpecialCharTok{:}\DecValTok{8554}\NormalTok{] }\OtherTok{\textless{}{-}} \StringTok{"NSP3"}
\NormalTok{name\_map[}\DecValTok{8555}\SpecialCharTok{:}\DecValTok{10054}\NormalTok{] }\OtherTok{\textless{}{-}} \StringTok{"NSP4"}
\NormalTok{name\_map[}\DecValTok{10055}\SpecialCharTok{:}\DecValTok{10972}\NormalTok{] }\OtherTok{\textless{}{-}} \StringTok{"3CL"}
\NormalTok{name\_map[}\DecValTok{10973}\SpecialCharTok{:}\DecValTok{11842}\NormalTok{] }\OtherTok{\textless{}{-}} \StringTok{"NSP6"}
\NormalTok{name\_map[}\DecValTok{11843}\SpecialCharTok{:}\DecValTok{12091}\NormalTok{] }\OtherTok{\textless{}{-}} \StringTok{"NSP7"}
\NormalTok{name\_map[}\DecValTok{12092}\SpecialCharTok{:}\DecValTok{12685}\NormalTok{] }\OtherTok{\textless{}{-}} \StringTok{"NSP8"}
\NormalTok{name\_map[}\DecValTok{12686}\SpecialCharTok{:}\DecValTok{13024}\NormalTok{] }\OtherTok{\textless{}{-}} \StringTok{"NSP9"}
\NormalTok{name\_map[}\DecValTok{13025}\SpecialCharTok{:}\DecValTok{13441}\NormalTok{] }\OtherTok{\textless{}{-}} \StringTok{"NSP10"}
\NormalTok{name\_map[}\DecValTok{13442}\SpecialCharTok{:}\DecValTok{16236}\NormalTok{] }\OtherTok{\textless{}{-}} \StringTok{"RdRp"}
\NormalTok{name\_map[}\DecValTok{16237}\SpecialCharTok{:}\DecValTok{18039}\NormalTok{] }\OtherTok{\textless{}{-}} \StringTok{"Helicase"}
\NormalTok{name\_map[}\DecValTok{18040}\SpecialCharTok{:}\DecValTok{19620}\NormalTok{] }\OtherTok{\textless{}{-}} \StringTok{"Exonuclease"}
\NormalTok{name\_map[}\DecValTok{19621}\SpecialCharTok{:}\DecValTok{20658}\NormalTok{] }\OtherTok{\textless{}{-}} \StringTok{"endoRNAse"}
\NormalTok{name\_map[}\DecValTok{20659}\SpecialCharTok{:}\DecValTok{21552}\NormalTok{] }\OtherTok{\textless{}{-}} \StringTok{"2’{-}O{-}ribose MTases"}
\NormalTok{name\_map[}\DecValTok{13442}\SpecialCharTok{:}\DecValTok{13480}\NormalTok{] }\OtherTok{\textless{}{-}} \StringTok{"NSP11"}
\NormalTok{name\_map[}\DecValTok{21563}\SpecialCharTok{:}\DecValTok{25384}\NormalTok{] }\OtherTok{\textless{}{-}} \StringTok{"Spike"}
\NormalTok{name\_map[}\DecValTok{25393}\SpecialCharTok{:}\DecValTok{26220}\NormalTok{] }\OtherTok{\textless{}{-}} \StringTok{"ORF3a"}
\NormalTok{name\_map[}\DecValTok{26245}\SpecialCharTok{:}\DecValTok{26472}\NormalTok{] }\OtherTok{\textless{}{-}} \StringTok{"Envelope"}
\NormalTok{name\_map[}\DecValTok{26523}\SpecialCharTok{:}\DecValTok{27191}\NormalTok{] }\OtherTok{\textless{}{-}} \StringTok{"Membrane"}
\NormalTok{name\_map[}\DecValTok{27202}\SpecialCharTok{:}\DecValTok{27387}\NormalTok{] }\OtherTok{\textless{}{-}} \StringTok{"ORF6"}
\NormalTok{name\_map[}\DecValTok{27394}\SpecialCharTok{:}\DecValTok{27759}\NormalTok{] }\OtherTok{\textless{}{-}} \StringTok{"ORF7a"}
\NormalTok{name\_map[}\DecValTok{27756}\SpecialCharTok{:}\DecValTok{27887}\NormalTok{] }\OtherTok{\textless{}{-}} \StringTok{"ORF7b"}
\NormalTok{name\_map[}\DecValTok{27894}\SpecialCharTok{:}\DecValTok{28259}\NormalTok{] }\OtherTok{\textless{}{-}} \StringTok{"ORF8"}
\NormalTok{name\_map[}\DecValTok{28274}\SpecialCharTok{:}\DecValTok{29533}\NormalTok{] }\OtherTok{\textless{}{-}} \StringTok{"Nucleocapsid"}
\NormalTok{name\_map[}\DecValTok{29558}\SpecialCharTok{:}\DecValTok{29674}\NormalTok{] }\OtherTok{\textless{}{-}} \StringTok{"ORF10"}
\NormalTok{mutation\_summaries\_combined}\SpecialCharTok{$}\NormalTok{protein }\OtherTok{\textless{}{-}}\NormalTok{ name\_map[mutation\_summaries\_combined}\SpecialCharTok{$}\NormalTok{position]}

\CommentTok{\# Remove rows with default protein name}
\NormalTok{mutation\_summaries\_combined }\OtherTok{\textless{}{-}} \FunctionTok{subset}\NormalTok{(mutation\_summaries\_combined, protein }\SpecialCharTok{!=} \StringTok{"Others"}\NormalTok{)}
\CommentTok{\# Color values adapted from settings used in original paper (Python code):}
\CommentTok{\# color\_continuous\_scale=[ \textquotesingle{}rgb(160,200,255)\textquotesingle{},\textquotesingle{}yellow\textquotesingle{},\textquotesingle{}rgba(13,143,129,0.7)\textquotesingle{},}
\CommentTok{\# \textquotesingle{}rgba(119,74,1750,0.8)\textquotesingle{}, \textquotesingle{}rgba(235,74,64,0.8)\textquotesingle{},\textquotesingle{}rgb(62,109,178)\textquotesingle{},}
\CommentTok{\# \textquotesingle{}rgb(160,200,255)\textquotesingle{}, \textquotesingle{}rgb(253,174,107)\textquotesingle{}, \textquotesingle{}rgba(119,74,1750,0.8)\textquotesingle{},}
\CommentTok{\# \textquotesingle{}rgb(160,200,255)\textquotesingle{},\textquotesingle{}black\textquotesingle{},\textquotesingle{}rgba(13,143,129,0.7)\textquotesingle{}, \textquotesingle{}rgba(119,74,1750,0.8)\textquotesingle{},}
\CommentTok{\# \textquotesingle{}rgba(235,74,64,0.8)\textquotesingle{},\textquotesingle{}rgb(62,109,178)\textquotesingle{}, \textquotesingle{}rgb(160,200,255)\textquotesingle{}, \textquotesingle{}black\textquotesingle{},}
\CommentTok{\# \textquotesingle{}rgb(260,200,255)\textquotesingle{}, \textquotesingle{}rgb(160,200,255)\textquotesingle{},\textquotesingle{}yellow\textquotesingle{},\textquotesingle{}rgba(13,143,129,0.7)\textquotesingle{},}
\CommentTok{\# \textquotesingle{}rgba(119,74,1750,0.8)\textquotesingle{}, \textquotesingle{}rgba(235,74,64,0.8)\textquotesingle{},\textquotesingle{}rgb(62,109,178)\textquotesingle{},}
\CommentTok{\# \textquotesingle{}rgb(160,200,255)\textquotesingle{}, \textquotesingle{}rgb(253,174,107)\textquotesingle{}])}
\NormalTok{scatter\_palette }\OtherTok{\textless{}{-}} \FunctionTok{c}\NormalTok{(}\StringTok{"\#a0c8ff"}\NormalTok{, }\StringTok{"yellow"}\NormalTok{, }\StringTok{"\#0d8f81"}\NormalTok{, }\StringTok{"\#774ad6"}\NormalTok{, }\StringTok{"\#eb4a40"}\NormalTok{, }\StringTok{"\#3e6db2"}\NormalTok{,}
    \StringTok{"\#a0c8ff"}\NormalTok{, }\StringTok{"\#fdae6b"}\NormalTok{, }\StringTok{"\#774ad6"}\NormalTok{, }\StringTok{"\#a0c8ff"}\NormalTok{, }\StringTok{"black"}\NormalTok{, }\StringTok{"\#0d8f81"}\NormalTok{, }\StringTok{"\#774ad6"}\NormalTok{, }\StringTok{"\#eb4a40"}\NormalTok{,}
    \StringTok{"\#3e6db2"}\NormalTok{, }\StringTok{"\#a0c8ff"}\NormalTok{, }\StringTok{"black"}\NormalTok{, }\StringTok{"\#04c8ff"}\NormalTok{, }\StringTok{"\#a0c8ff"}\NormalTok{, }\StringTok{"yellow"}\NormalTok{, }\StringTok{"\#0d8f81"}\NormalTok{, }\StringTok{"\#774ad6"}\NormalTok{,}
    \StringTok{"\#eb4a40"}\NormalTok{, }\StringTok{"\#3e6db2"}\NormalTok{, }\StringTok{"\#a0c8ff"}\NormalTok{, }\StringTok{"\#fdae6b"}\NormalTok{)}

\CommentTok{\# Create scatter plot of mutations}
\NormalTok{scatter\_fig }\OtherTok{\textless{}{-}} \FunctionTok{plot\_ly}\NormalTok{(}\AttributeTok{data =}\NormalTok{ mutation\_summaries\_combined, }\AttributeTok{x =} \SpecialCharTok{\textasciitilde{}}\NormalTok{position, }\AttributeTok{y =} \SpecialCharTok{\textasciitilde{}}\FunctionTok{log}\NormalTok{(frequency),}
    \AttributeTok{type =} \StringTok{"scatter"}\NormalTok{, }\AttributeTok{mode =} \StringTok{"markers"}\NormalTok{, }\AttributeTok{color =} \SpecialCharTok{\textasciitilde{}}\NormalTok{protein, }\AttributeTok{colors =}\NormalTok{ scatter\_palette)}
\NormalTok{scatter\_fig }\OtherTok{\textless{}{-}}\NormalTok{ scatter\_fig }\SpecialCharTok{\%\textgreater{}\%}
    \FunctionTok{layout}\NormalTok{(}\AttributeTok{title =} \StringTok{"Mutation distribution across the SARS{-}CoV{-}2 genome"}\NormalTok{, }\AttributeTok{xaxis =} \FunctionTok{list}\NormalTok{(}\AttributeTok{title =} \StringTok{""}\NormalTok{,}
        \AttributeTok{showgrid =} \ConstantTok{FALSE}\NormalTok{, }\AttributeTok{tickangle =} \DecValTok{60}\NormalTok{, }\AttributeTok{tickmode =} \StringTok{"array"}\NormalTok{, }\AttributeTok{nticks =} \DecValTok{11}\NormalTok{, }\AttributeTok{tickvals =} \FunctionTok{c}\NormalTok{(}\DecValTok{536}\NormalTok{,}
            \DecValTok{5638}\NormalTok{, }\DecValTok{9305}\NormalTok{, }\DecValTok{10514}\NormalTok{, }\DecValTok{11408}\NormalTok{, }\DecValTok{14839}\NormalTok{, }\DecValTok{17138}\NormalTok{, }\DecValTok{23474}\NormalTok{, }\DecValTok{25807}\NormalTok{, }\DecValTok{26359}\NormalTok{, }\DecValTok{28904}\NormalTok{),}
        \AttributeTok{ticktext =} \FunctionTok{c}\NormalTok{(}\StringTok{"NSP1"}\NormalTok{, }\StringTok{"NSP3"}\NormalTok{, }\StringTok{"NSP4"}\NormalTok{, }\StringTok{"3CL"}\NormalTok{, }\StringTok{"NSP6"}\NormalTok{, }\StringTok{"RdRp"}\NormalTok{, }\StringTok{"Helicase"}\NormalTok{, }\StringTok{"S"}\NormalTok{,}
            \StringTok{"ORF3a"}\NormalTok{, }\StringTok{"E"}\NormalTok{, }\StringTok{"N"}\NormalTok{)), }\AttributeTok{yaxis =} \FunctionTok{list}\NormalTok{(}\AttributeTok{title =} \StringTok{"ln(Frequency)"}\NormalTok{, }\AttributeTok{showgrid =} \ConstantTok{FALSE}\NormalTok{),}
        \AttributeTok{font =} \FunctionTok{list}\NormalTok{(}\AttributeTok{family =} \StringTok{"Serif"}\NormalTok{, }\AttributeTok{size =} \DecValTok{32}\NormalTok{, }\AttributeTok{color =} \StringTok{"black"}\NormalTok{), }\AttributeTok{showlegend =} \ConstantTok{FALSE}\NormalTok{,}
        \AttributeTok{margin =} \FunctionTok{list}\NormalTok{(}\AttributeTok{t =} \DecValTok{200}\NormalTok{))}
\FunctionTok{save\_image}\NormalTok{(scatter\_fig, }\StringTok{"./figures/updated\_distribution.pdf"}\NormalTok{, }\AttributeTok{width =} \DecValTok{2304}\NormalTok{, }\AttributeTok{height =} \DecValTok{1152}\NormalTok{)}
\NormalTok{knitr}\SpecialCharTok{::}\FunctionTok{include\_graphics}\NormalTok{(}\StringTok{"./figures/updated\_distribution.pdf"}\NormalTok{)}
\CommentTok{\# Generate mutation statistics by protein}
\NormalTok{aggregate\_summaries }\OtherTok{\textless{}{-}}\NormalTok{ mutation\_summaries\_combined }\SpecialCharTok{\%\textgreater{}\%}
    \FunctionTok{group\_by}\NormalTok{(protein) }\SpecialCharTok{\%\textgreater{}\%}
    \FunctionTok{summarise}\NormalTok{(}\AttributeTok{total\_mutations =} \FunctionTok{sum}\NormalTok{(frequency)) }\SpecialCharTok{\%\textgreater{}\%}
    \FunctionTok{mutate}\NormalTok{(}\AttributeTok{raw\_percent =}\NormalTok{ (total\_mutations}\SpecialCharTok{/}\FunctionTok{sum}\NormalTok{(total\_mutations)) }\SpecialCharTok{*} \DecValTok{100}\NormalTok{, }\AttributeTok{nucleotide\_length =} \FunctionTok{sapply}\NormalTok{(protein,}
        \ControlFlowTok{switch}\NormalTok{, }\AttributeTok{NSP1 =}\NormalTok{ \{}
            \DecValTok{805} \SpecialCharTok{{-}} \DecValTok{266}
\NormalTok{        \}, }\AttributeTok{NSP2 =}\NormalTok{ \{}
            \DecValTok{2719} \SpecialCharTok{{-}} \DecValTok{806}
\NormalTok{        \}, }\AttributeTok{NSP3 =}\NormalTok{ \{}
            \DecValTok{8554} \SpecialCharTok{{-}} \DecValTok{2720}
\NormalTok{        \}, }\AttributeTok{NSP4 =}\NormalTok{ \{}
            \DecValTok{10054} \SpecialCharTok{{-}} \DecValTok{8555}
\NormalTok{        \}, }\StringTok{\textasciigrave{}}\AttributeTok{3CL}\StringTok{\textasciigrave{}} \OtherTok{=}\NormalTok{ \{}
            \DecValTok{10972} \SpecialCharTok{{-}} \DecValTok{10055}
\NormalTok{        \}, }\AttributeTok{NSP6 =}\NormalTok{ \{}
            \DecValTok{11842} \SpecialCharTok{{-}} \DecValTok{10973}
\NormalTok{        \}, }\AttributeTok{NSP7 =}\NormalTok{ \{}
            \DecValTok{12091} \SpecialCharTok{{-}} \DecValTok{11843}
\NormalTok{        \}, }\AttributeTok{NSP8 =}\NormalTok{ \{}
            \DecValTok{12685} \SpecialCharTok{{-}} \DecValTok{12092}
\NormalTok{        \}, }\AttributeTok{NSP9 =}\NormalTok{ \{}
            \DecValTok{13024} \SpecialCharTok{{-}} \DecValTok{12686}
\NormalTok{        \}, }\AttributeTok{NSP10 =}\NormalTok{ \{}
            \DecValTok{13441} \SpecialCharTok{{-}} \DecValTok{13025}
\NormalTok{        \}, }\AttributeTok{RdRp =}\NormalTok{ \{}
            \DecValTok{16236} \SpecialCharTok{{-}} \DecValTok{13442}
\NormalTok{        \}, }\AttributeTok{Helicase =}\NormalTok{ \{}
            \DecValTok{18039} \SpecialCharTok{{-}} \DecValTok{16237}
\NormalTok{        \}, }\AttributeTok{Exonuclease =}\NormalTok{ \{}
            \DecValTok{19620} \SpecialCharTok{{-}} \DecValTok{18040}
\NormalTok{        \}, }\AttributeTok{endoRNAse =}\NormalTok{ \{}
            \DecValTok{20658} \SpecialCharTok{{-}} \DecValTok{19621}
\NormalTok{        \}, }\StringTok{\textasciigrave{}}\AttributeTok{2’{-}O{-}ribose MTases}\StringTok{\textasciigrave{}} \OtherTok{=}\NormalTok{ \{}
            \DecValTok{21552} \SpecialCharTok{{-}} \DecValTok{20659}
\NormalTok{        \}, }\AttributeTok{NSP11 =}\NormalTok{ \{}
            \DecValTok{13480} \SpecialCharTok{{-}} \DecValTok{13442}
\NormalTok{        \}, }\AttributeTok{Spike =}\NormalTok{ \{}
            \DecValTok{25384} \SpecialCharTok{{-}} \DecValTok{21563}
\NormalTok{        \}, }\AttributeTok{ORF3a =}\NormalTok{ \{}
            \DecValTok{26220} \SpecialCharTok{{-}} \DecValTok{25393}
\NormalTok{        \}, }\AttributeTok{Envelope =}\NormalTok{ \{}
            \DecValTok{26472} \SpecialCharTok{{-}} \DecValTok{26245}
\NormalTok{        \}, }\AttributeTok{Membrane =}\NormalTok{ \{}
            \DecValTok{27191} \SpecialCharTok{{-}} \DecValTok{26523}
\NormalTok{        \}, }\AttributeTok{ORF6 =}\NormalTok{ \{}
            \DecValTok{27387} \SpecialCharTok{{-}} \DecValTok{27202}
\NormalTok{        \}, }\AttributeTok{ORF7a =}\NormalTok{ \{}
            \DecValTok{27759} \SpecialCharTok{{-}} \DecValTok{27394}
\NormalTok{        \}, }\AttributeTok{ORF7b =}\NormalTok{ \{}
            \DecValTok{27887} \SpecialCharTok{{-}} \DecValTok{27756}
\NormalTok{        \}, }\AttributeTok{ORF8 =}\NormalTok{ \{}
            \DecValTok{28259} \SpecialCharTok{{-}} \DecValTok{27894}
\NormalTok{        \}, }\AttributeTok{Nucleocapsid =}\NormalTok{ \{}
            \DecValTok{29533} \SpecialCharTok{{-}} \DecValTok{28274}
\NormalTok{        \}, }\AttributeTok{ORF10 =}\NormalTok{ \{}
            \DecValTok{29674} \SpecialCharTok{{-}} \DecValTok{29558}
\NormalTok{        \}), }\AttributeTok{mutations\_per\_nucleotide =}\NormalTok{ total\_mutations}\SpecialCharTok{/}\NormalTok{nucleotide\_length) }\SpecialCharTok{\%\textgreater{}\%}
    \FunctionTok{arrange}\NormalTok{(}\FunctionTok{desc}\NormalTok{(total\_mutations))}
\NormalTok{knitr}\SpecialCharTok{::}\FunctionTok{kable}\NormalTok{(aggregate\_summaries, }\AttributeTok{format =} \StringTok{"latex"}\NormalTok{, }\AttributeTok{booktabs =} \ConstantTok{TRUE}\NormalTok{, }\AttributeTok{digits =} \DecValTok{3}\NormalTok{,}
    \AttributeTok{col.names =} \FunctionTok{c}\NormalTok{(}\StringTok{"protein"}\NormalTok{, }\StringTok{"total mutations"}\NormalTok{, }\StringTok{"\% total mutations"}\NormalTok{, }\StringTok{"protein length"}\NormalTok{,}
        \StringTok{"mutations/nucleotide"}\NormalTok{), }\AttributeTok{caption =} \StringTok{"Summary statistics for each SARS{-}CoV{-}2 protein (author W.S.)"}\NormalTok{,}
    \AttributeTok{format.args =} \FunctionTok{list}\NormalTok{(}\AttributeTok{big.mark =} \StringTok{","}\NormalTok{))}
\CommentTok{\# Read mutation summary data for 10202021}
\NormalTok{path }\OtherTok{\textless{}{-}} \StringTok{"./data/MutationSummary\_10202021/"}
\NormalTok{files }\OtherTok{\textless{}{-}} \FunctionTok{dir}\NormalTok{(path)}
\NormalTok{mutation\_summaries\_10202021 }\OtherTok{\textless{}{-}} \FunctionTok{do.call}\NormalTok{(rbind, }\FunctionTok{lapply}\NormalTok{(}\FunctionTok{paste0}\NormalTok{(path, files), read.csv))}

\CommentTok{\# Find percent mutations by protein for 10202021}
\NormalTok{bar\_data\_10202021 }\OtherTok{\textless{}{-}}\NormalTok{ mutation\_summaries\_10202021 }\SpecialCharTok{\%\textgreater{}\%}
    \FunctionTok{mutate}\NormalTok{(}\AttributeTok{position =} \FunctionTok{as.integer}\NormalTok{(}\FunctionTok{gsub}\NormalTok{(}\StringTok{"[\^{}0{-}9]"}\NormalTok{, }\StringTok{""}\NormalTok{, mutation\_site))) }\SpecialCharTok{\%\textgreater{}\%}
    \FunctionTok{group\_by}\NormalTok{(position) }\SpecialCharTok{\%\textgreater{}\%}
    \FunctionTok{summarise}\NormalTok{(}\AttributeTok{frequency =} \FunctionTok{sum}\NormalTok{(Total\_frequency))}
\NormalTok{bar\_data\_10202021}\SpecialCharTok{$}\NormalTok{protein }\OtherTok{\textless{}{-}}\NormalTok{ name\_map[bar\_data\_10202021}\SpecialCharTok{$}\NormalTok{position]}
\NormalTok{bar\_data\_10202021 }\OtherTok{\textless{}{-}} \FunctionTok{subset}\NormalTok{(bar\_data\_10202021, protein }\SpecialCharTok{!=} \StringTok{"Others"}\NormalTok{)}
\NormalTok{bar\_data\_10202021 }\OtherTok{\textless{}{-}}\NormalTok{ bar\_data\_10202021 }\SpecialCharTok{\%\textgreater{}\%}
    \FunctionTok{group\_by}\NormalTok{(protein) }\SpecialCharTok{\%\textgreater{}\%}
    \FunctionTok{summarise}\NormalTok{(}\AttributeTok{total\_mutations =} \FunctionTok{sum}\NormalTok{(frequency)) }\SpecialCharTok{\%\textgreater{}\%}
    \FunctionTok{mutate}\NormalTok{(}\AttributeTok{raw\_percent =}\NormalTok{ (total\_mutations}\SpecialCharTok{/}\FunctionTok{sum}\NormalTok{(total\_mutations)) }\SpecialCharTok{*} \DecValTok{100}\NormalTok{) }\SpecialCharTok{\%\textgreater{}\%}
    \FunctionTok{arrange}\NormalTok{(}\FunctionTok{desc}\NormalTok{(total\_mutations)) }\SpecialCharTok{\%\textgreater{}\%}
    \FunctionTok{select}\NormalTok{(protein, raw\_percent) }\SpecialCharTok{\%\textgreater{}\%}
    \FunctionTok{head}\NormalTok{(}\DecValTok{7}\NormalTok{)}
\NormalTok{others }\OtherTok{\textless{}{-}} \FunctionTok{list}\NormalTok{(}\StringTok{"Others"}\NormalTok{, }\DecValTok{100} \SpecialCharTok{{-}} \FunctionTok{sum}\NormalTok{(bar\_data\_10202021}\SpecialCharTok{$}\NormalTok{raw\_percent))}
\NormalTok{bar\_data\_10202021 }\OtherTok{\textless{}{-}}\NormalTok{ bar\_data\_10202021 }\SpecialCharTok{\%\textgreater{}\%}
    \FunctionTok{rbind}\NormalTok{(others) }\SpecialCharTok{\%\textgreater{}\%}
    \FunctionTok{arrange}\NormalTok{(raw\_percent) }\SpecialCharTok{\%\textgreater{}\%}
    \FunctionTok{mutate}\NormalTok{(}\AttributeTok{date =} \StringTok{"10202021"}\NormalTok{)}

\CommentTok{\# Find percent mutations by protein for latest data (10312021)}
\NormalTok{bar\_data\_10312021 }\OtherTok{\textless{}{-}}\NormalTok{ aggregate\_summaries }\SpecialCharTok{\%\textgreater{}\%}
    \FunctionTok{select}\NormalTok{(protein, raw\_percent) }\SpecialCharTok{\%\textgreater{}\%}
    \FunctionTok{head}\NormalTok{(}\DecValTok{7}\NormalTok{)}
\NormalTok{others }\OtherTok{\textless{}{-}} \FunctionTok{list}\NormalTok{(}\StringTok{"Others"}\NormalTok{, }\DecValTok{100} \SpecialCharTok{{-}} \FunctionTok{sum}\NormalTok{(bar\_data\_10312021}\SpecialCharTok{$}\NormalTok{raw\_percent))}
\NormalTok{bar\_data\_10312021 }\OtherTok{\textless{}{-}}\NormalTok{ bar\_data\_10312021 }\SpecialCharTok{\%\textgreater{}\%}
    \FunctionTok{rbind}\NormalTok{(others) }\SpecialCharTok{\%\textgreater{}\%}
    \FunctionTok{arrange}\NormalTok{(raw\_percent) }\SpecialCharTok{\%\textgreater{}\%}
    \FunctionTok{mutate}\NormalTok{(}\AttributeTok{date =} \StringTok{"10312021"}\NormalTok{)}

\CommentTok{\# Combine data into single data frame for plotting}
\NormalTok{bar\_data }\OtherTok{\textless{}{-}} \FunctionTok{rbind}\NormalTok{(bar\_data\_10202021, bar\_data\_10312021)}

\CommentTok{\# Created stacked bar chart for each date ordered by percent}
\NormalTok{stacked\_bar }\OtherTok{\textless{}{-}} \FunctionTok{ggplot}\NormalTok{(bar\_data, }\FunctionTok{aes}\NormalTok{(}\AttributeTok{x =}\NormalTok{ date, }\AttributeTok{y =}\NormalTok{ raw\_percent, }\AttributeTok{group =}\NormalTok{ raw\_percent)) }\SpecialCharTok{+}
    \FunctionTok{geom\_col}\NormalTok{(}\FunctionTok{aes}\NormalTok{(}\AttributeTok{fill =}\NormalTok{ protein), }\AttributeTok{width =} \FloatTok{0.9}\NormalTok{) }\SpecialCharTok{+} \FunctionTok{scale\_fill\_brewer}\NormalTok{(}\AttributeTok{type =} \StringTok{"qual"}\NormalTok{,}
    \AttributeTok{palette =} \StringTok{"Paired"}\NormalTok{) }\SpecialCharTok{+} \FunctionTok{geom\_text}\NormalTok{(}\FunctionTok{aes}\NormalTok{(}\AttributeTok{label =} \FunctionTok{paste0}\NormalTok{(protein, }\StringTok{" {-} "}\NormalTok{, }\FunctionTok{round}\NormalTok{(raw\_percent,}
    \AttributeTok{digits =} \DecValTok{1}\NormalTok{), }\StringTok{"\%"}\NormalTok{)), }\AttributeTok{position =} \FunctionTok{position\_stack}\NormalTok{(}\AttributeTok{vjust =} \FloatTok{0.5}\NormalTok{)) }\SpecialCharTok{+} \FunctionTok{ggtitle}\NormalTok{(}\StringTok{"Percentage of total mutations by protein"}\NormalTok{) }\SpecialCharTok{+}
    \FunctionTok{xlab}\NormalTok{(}\StringTok{"Date"}\NormalTok{) }\SpecialCharTok{+} \FunctionTok{ylab}\NormalTok{(}\StringTok{"Percent"}\NormalTok{) }\SpecialCharTok{+} \FunctionTok{labs}\NormalTok{(}\AttributeTok{fill =} \StringTok{"Protein"}\NormalTok{) }\SpecialCharTok{+} \FunctionTok{theme\_classic}\NormalTok{(}\AttributeTok{base\_family =} \StringTok{"serif"}\NormalTok{) }\SpecialCharTok{+}
    \FunctionTok{theme}\NormalTok{(}\AttributeTok{plot.title =} \FunctionTok{element\_text}\NormalTok{(}\AttributeTok{hjust =} \FloatTok{0.5}\NormalTok{))}
\FunctionTok{ggsave}\NormalTok{(}\StringTok{"./figures/stacked\_bar.pdf"}\NormalTok{)}
\NormalTok{knitr}\SpecialCharTok{::}\FunctionTok{include\_graphics}\NormalTok{(}\StringTok{"./figures/stacked\_bar.pdf"}\NormalTok{)}
\CommentTok{\# Cite packages}
\NormalTok{knitr}\SpecialCharTok{::}\FunctionTok{write\_bib}\NormalTok{(}\FunctionTok{c}\NormalTok{(}\FunctionTok{.packages}\NormalTok{(), }\StringTok{"bookdown"}\NormalTok{), }\StringTok{"packages.bib"}\NormalTok{)}
\FunctionTok{sessionInfo}\NormalTok{(}\AttributeTok{package =} \ConstantTok{NULL}\NormalTok{)}
\end{Highlighting}
\end{Shaded}

\hypertarget{session-info}{%
\section{Session Info}\label{session-info}}

\begin{verbatim}
## R version 4.1.1 (2021-08-10)
## Platform: x86_64-apple-darwin17.0 (64-bit)
## Running under: macOS Big Sur 10.16
## 
## Matrix products: default
## BLAS:   /Library/Frameworks/R.framework/Versions/4.1/Resources/lib/libRblas.0.dylib
## LAPACK: /Library/Frameworks/R.framework/Versions/4.1/Resources/lib/libRlapack.dylib
## 
## locale:
## [1] en_US.UTF-8/en_US.UTF-8/en_US.UTF-8/C/en_US.UTF-8/en_US.UTF-8
## 
## attached base packages:
## [1] stats     graphics  grDevices utils     datasets  methods   base     
## 
## other attached packages:
## [1] bookdown_0.24   plotly_4.10.0   ggplot2_3.3.5   reticulate_1.22
## [5] dplyr_1.0.7     rticles_0.21    formatR_1.11    knitr_1.36     
## 
## loaded via a namespace (and not attached):
##  [1] tidyselect_1.1.1   xfun_0.27          purrr_0.3.4        lattice_0.20-45   
##  [5] colorspace_2.0-2   vctrs_0.3.8        generics_0.1.0     htmltools_0.5.2   
##  [9] viridisLite_0.4.0  yaml_2.2.1         utf8_1.2.2         rlang_0.4.12      
## [13] pillar_1.6.4       glue_1.4.2         withr_2.4.2        DBI_1.1.1         
## [17] RColorBrewer_1.1-2 lifecycle_1.0.1    stringr_1.4.0      munsell_0.5.0     
## [21] gtable_0.3.0       htmlwidgets_1.5.4  codetools_0.2-18   evaluate_0.14     
## [25] labeling_0.4.2     fastmap_1.1.0      crosstalk_1.2.0    fansi_0.5.0       
## [29] Rcpp_1.0.7         scales_1.1.1       jsonlite_1.7.2     farver_2.1.0      
## [33] png_0.1-7          digest_0.6.28      stringi_1.7.5      grid_4.1.1        
## [37] rprojroot_2.0.2    here_1.0.1         tools_4.1.1        magrittr_2.0.1    
## [41] lazyeval_0.2.2     tibble_3.1.5       crayon_1.4.1       tidyr_1.1.4       
## [45] pkgconfig_2.0.3    ellipsis_0.3.2     Matrix_1.3-4       data.table_1.14.2 
## [49] assertthat_0.2.1   rmarkdown_2.11     httr_1.4.2         R6_2.5.1          
## [53] compiler_4.1.1
\end{verbatim}

\bibliographystyle{unsrt}
\bibliography{packages.bibreferences.bib}

\end{document}